# Pressure Effects on the Atomic and Electronic Structure of Aligned Small Diameter Carbon Nanotubes

*Sumit Saxena\*, Trevor A. Tyson[†]*

[\*†]Department of Physics, New Jersey Institute of Technology, University heights, Newark, NJ 07102-1982

\*ss524@njit.edu, [†] tyson@adm.njit.edu

ABSTRACT.

Density functional methods have been used to calculate the electronic properties of aligned small diameter single-walled carbon nanotubes under hydrostatic pressures. Abrupt pressure induced semiconductor-metal and metal-semiconductor transitions concomitant with changes in structure are observed. Novel and unexpected unit cell nanotube cross-sections are found. These tubes are observed to form interlinking structures at very high pressures. The large changes in electronic structure and the ability of different small diameter nanotubes to respond in different pressure regimes suggest their use in high pressure multiple switching devices.

# I. Introduction

The ability to control transport properties using mechanical deformations will boost the development of nano electromechanical devices[1,2,3,4,5]. Strain produces structural deformations in carbon nanotubes, causing change in electronic structure[6,7,8,9,10,11,12]. Carbon nanotubes have high stiffness and large axial strength[13]. While all studies have focused on pressure effects on large diameter nanotubes which are unsuitable for high pressure applications, small diameter nanotubes still remain unexplored. Individual single walled carbon nanotubes (SWCNT's) when subjected to strain exhibit shifts in the vibrational frequencies of the radial breathing mode and G mode. These shifts are due to the structural changes. Reversible increase in Raman intensities is also observed indicating modification of the band structure[14]. We report for the first time unique structural and electronic properties of small diameter zigzag single-walled carbon nanotubes under hydrostatic pressure, using density functional theory. The calculations are performed for aligned tubes. These nanotubes are linked at high pressures.

In this work, we provide a systematic study of the effect of hydrostatic pressure on the electronic structure of small diameter zigzag (n, 0) SWCNT's $6 \leq n \leq 9$. We focus mainly on the evolution of the density of states (DOS) with pressure and attempt to understand the structural transformations concomitant with changes in electronic properties under high pressures. We found that the structural changes are abrupt and produce large changes in the corresponding electronic structure. At high pressures the large changes in cross-section have been found to be related to interaction between neighboring tubes. In similar calculations large diameter nanotubes have been reported in literature to buckle at the center forming dumbbell shaped cross-sections. This is completely different to the type of buckling which we have observed in small diameter nanotubes. At high pressures depending on the tube chiralities, it is observed that the tubes form interlinked three dimensional structures.

## II. Computational Methods

First principles calculations were carried out using the plane wave basis Vienna Ab Initio Simulation Package[15, 16] (VASP) implementing the spin restricted density functional theory (DFT) in the local density approximation (LDA). Highly accurate projected augmented wave[17] (PAW) potentials using PBE[18] exchange correlation have been used as reported in earlier studies on carbon nanotubes[19, 20, 21]. A kinetic energy cutoff of 550 eV has been used to describe the electronic potentials. The geometry was relaxed using the conjugate gradient technique with 6 K points along the tube axis; such that no forces on the atoms exceed 0.001 eV/Å. Hydrostatic pressure is monotonically increased in steps of 1 GPa by equating the diagonal elements of the stress tensor to the desired pressure. Orthorhombic super-cells were constructed, such that the separation between the adjacent nanotubes was about 8Å in the radial direction in-order to reduce interactions. A unit cell consisting of 4 zigzag carbon atom rings with length of about 8.54Å along the tube axis was used to model an infinitely long tube without end-caps by utilizing periodic boundary conditions.

## III. Results and Discussion

Pressure induced metallization can be understood on the basis of increased inter-atomic interactions at certain sites due to breaking of the circular symmetry of the cross-section and enhanced $\sigma^*$ - $\pi^*$ hybridization that lead to electronic state overlap and closure of the band gap. The cross sections of the tube can be observed in the charge density plots Fig. 1(a) to Fig. 1(c) for (7, 0) SWCNT, Fig. 2(a) to Fig. 2(c) for (8, 0) SWCNT, Fig. 3(a) to Fig. 3(c) for (6, 0) SWCNT and Fig. 4(a) to Fig. 4(c) for (9, 0) SWCNT. These clearly show structural changes with increase of pressure forming novel unit cell cross-sections. At moderate pressures minimal interaction between the tubes is observed as low charge density along connecting line in the charge density plots.

In order to understand the pressure induced structure - transport correlations we simultaneously study the evolution of DOS at the Fermi level with pressure and analyze the corresponding changes in structure and bond lengths for metallic as well as semi-conducting tubes. The contour plots of the DOS of tubes in Fig. 1(d), Fig. 2(d), Fig. 3(d) and Fig. 4(d) reveals the changes in electronic structure with pressure. We will first focus our attention on tubes showing semiconducting behavior and then those showing metallic behavior at ambient pressures. We observe initial closing of the band gap in semi-conducting (7, 0) and (8, 0) nanotubes at moderate pressures in the upper one dimensional plots of the DOS at the Fermi level in Fig. 1(d) and Fig. 2(d) near 8 GPa and 20 GPa, respectively. These results are consistent with the experimental observations for large diameter nanotubes[22] where the band gap of semi-conducting tubes always decreases with pressure. Careful analysis of the structure revealed that as pressure is increased the tube starts to flatten gradually and some states/cell ($\sim 0.1 \ast 10^{-4}$) start to appear in the DOS at the Fermi level at pressures as low as 3 GPa. However, at about 7 GPa sudden flattening of the tube marked by a sudden change in the DOS at the Fermi level occurs as seen in Fig. 1(d) is observed. This indicates a definite detectable semiconductor – metal transition at pressures of about 7 GPa. It seems that as the pressure is increased further the charges begin to delocalize more along the curved edges of the elliptic cross-section as observed in the Fig. 1(a) and an increase in the DOS is observed. The tubes forms a very weakly interlinked structure at low pressure of 8 GPa as the interlinking bonds marked by white arrows and the intratube bonds length marked by black arrows are 1.55Å and 1.52Å respectively. As we increase the pressure an anomalous metallic to semiconducting phase transformation is observed. The tubes get more flattened as seen for the cross-sections at 20 GPa in Fig. 1(b). An overall decrease in the bond lengths in the structure is observed. The representative bonds marked by white and black arrows decrease in length to about 1.52Å and 1.50Å indicating increased talking between the tubes. As the pressure is further increased to 60 GPa the interlinking bonds indicated by white arrows and the intratube bonds represented by black arrows in Fig. 1(c) become equivalent and are 1.46Å. At this stage the smallest bond lengths are 1.38Å and all the bonds become comparable to regular C-C bond length of 1.42Å indicating the formation of an interlinked three dimensional structure.

The (8, 0) SWCNT is known to show semiconducting behavior under ambient conditions hence as the pressure is increased semiconductor – metallic transition is expected due to changes in curvature[12]. The sample cross-section at 8 GPa in Fig. 2(a) show that the tubes are non interacting. The transition pressure is expected to be higher than that required for the (7, 0) nanotube as it has a larger band gap. The band gap between the bottom of the conduction band and the top of the valence band at the Γ point is about 0.6 eV, this is in agreement with previous published results[23]. It is observed that the (8, 0) tube shows this behavior at about 20 GPa as observed in the plot of DOS at Fermi level in top inset in the Fig. 2(d). On further increase of pressure a change in the density of states is observed at about 30 GPa. The inset to the right in Fig. 2(d) shows that the tube remains metallic. The sample cross-section of charge densities in Fig. 2(b) indicates that at this point the interaction between the tubes can be neglected. Interaction with other nanotubes starts to show up only at pressures of ~38 GPa. This abrupt change in the DOS is indicative of another structural transition in which the nanotubes come so close to each other that they start interacting and form a three dimensional structure as seen in Fig. 2(d). The structure remains in a semi-conducting state. The light and dark arrows indicate regions with weak and strong bonds, respectively. The arrows have the same meaning in all charge density figures.

The (6, 0) and (9, 0) nanotubes have been predicted to be metallic by the (n-m) mod 3 rule. We observe that as pressure is increased on the (6, 0) nanotube it undergoes a metal – semiconductor transition at about 20 GPa as in Fig. 3(d). The density of states at the Fermi level reaches a maximum at 10 GPa. This is marked by decrease in bond lengths by 0.1% to 0.53% along different bonds at this pressure. A sample cross-section at 8 GPa in Fig. 3(a) indicates that no structural transformation occurs and that change in DOS is an indication of relative changes in the bond lengths. The cross section of the structure remain preserved up to about 20 GPa as in Fig. 3(b) and the tube remains metallic although the DOS at the Fermi level drops considerably as seen in the inset of DOS plot at 20 GPa in Fig. 3(d). It can be seen in Fig. 3(b) that the interactions between the tubes is still weak at 20 GPa. As pressure is further increased structural transformation lead to opening of band gap. This is indicated by a black rectangular band in the contour plot of the DOS at different pressures. For pressures above 25 GPa the band gap remains approximately constant. This is consistent with the fact that hardly any structural transformations occur as pressure is increased. The applied pressure only tends to bring the deformed tubes closer.

The (9, 0) tube does not show any metal – semiconductor transitions. However, the density of states in the inset of DOS plot at the Fermi level in Fig. 4(d) is observed to increase tremendously at low pressures of about 2 GPa and remains metallic over the entire pressure range up to 60 GPa. As the pressure is increased from 1 GPa to 2 Gpa, very small changes in cross-sections were observed. Thereafter no appreciable changes in the cross-section are observed upto 50 GPa as in Fig. 4(a) and Fig. 4(b), which are representative cross-sections at 8 GPa and 30 GPa. The bond lengths decrease along the more curved edges of the elliptic cross-sections and increase at the flattened surface where the atoms interact weakly. As the pressure is further increased no significant changes in the bond length are observed until ~50 GPa where a small bump in the DOS is observed the jump was found to be about 10%. This increase in the DOS occurs due to interlinking of the tubes as observed in the plot of charge density for (9, 0) tubes at 60 GPA. The interlinking bonds marked by white arrows are about 1.54 Å. The highly localized charge channels marked by the black arrows are about 1.33Å. The charge from these highly localized charge zones spread gradually starting from center towards the buckled ends, which shows up in form of decreased bond lengths of 1.43Å, 1.46Å, 1.51 Å and 1.50Å corresponding to unmarked bonds. The interlinking bonds marked by white arrows are the weakest with the bond length of 1.55Å. This picture of bond lengths suggests the formation of highly linked three dimensional structures possibly capable of transporting charges effectively.

## IV. Conclusions

In conclusion using first principle calculations, we have studied in detail the effect of hydrostatic pressure on small diameter SWCNT. The cross sections transform from circular to novel geometries at high pressures. Pressure induced modification of the band gap concomitant with abrupt changes in cross sections due to interlinking of the tubes at high pressures is observed. The band gap of small diameter zig-zag semi-conducting nanotubes decreases with hydrostatic pressure. A large number of pressure induced electronic phases are observed in semi-conducting nanotubes. These transitions correlates closely to the pressure induced structural transformation in bond lengths, bond angles and formation of new bonds. We finally note that the density of states at Fermi level of different small diameter nanotubes responds strongly to pressure, this suggests the potential for the development of pressure switches with multi switching capabilities over a broad pressure range.

# **Acknowledgement**

This work has been supported in part by NSF DMR grant DMR-0512196.

# Appendix A

The following figures show the calculated band structure of (6, 0), (7, 0), (8, 0) and (9, 0) carbon nanotubes along the tube axis at different critical pressures.

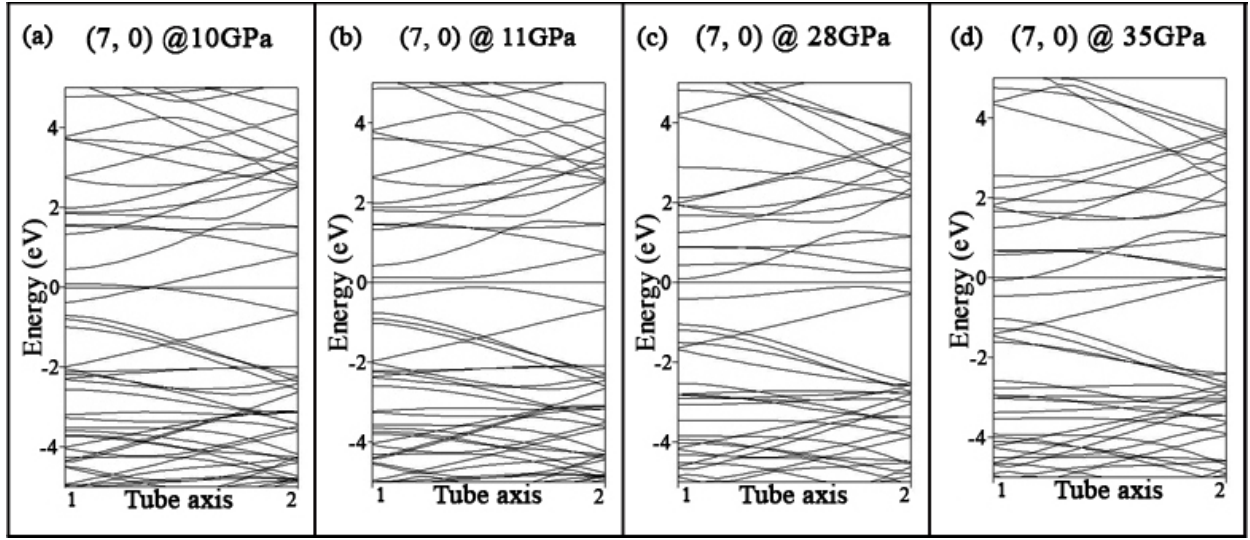

Fig.A1: - The plot of band structure diagram for semi-conducting (7, 0) SWCNT are shown above in Fig. (a) – Fig. (d) at pressures of 10 GPa, 11 GPa, 28 GPa and 35 GPa. The Fermi level is at 0 eV.

The crossing of bands at the Fermi level is observed in Fig.A1 (a). This indicates the metallic character of the tubes at this pressure. As the pressure is further increased to 11 GPa, a direct band-gap opens up the Fermi level as observed in Fig.A1 (b) above and the tube starts to behave as semiconductor. This is also observed in the one dimensional DOS plot inset in Fig. 1(d). As the pressure is increased further we see the formation of indirect band gap in Fig.A1 (c). This indicates that the tube behaves as a semiconductor at this pressure of 28 GPa. At about 30 GPa the band-gap starts to close down and due to electron smearing effects the tube begins to show a metallic behavior. The sample plot of band structure at 35 GPA in Fig.A1 (d) indicates crossing of states at the bottom of the conduction band and the upper states of the valence bands at the Fermi level. This shows that the carbon nanotubes becomes metallic and suggest that a jump in the DOS be expected around this pressure range.

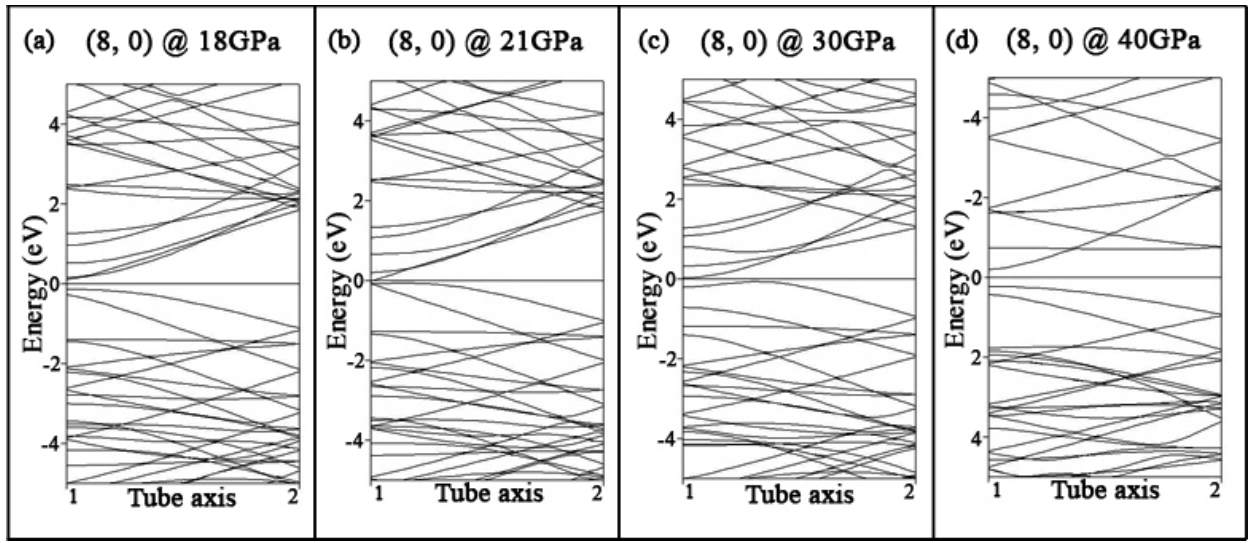

Fig.A2: - The plot of band structure diagram for semiconducting (8, 0) SWCNT are shown above in Fig. (a) – Fig.(d) at pressures of 18 GPa, 21 GPa, 30 GPa and 40 GPa. The Fermi level is at 0 eV.

We see a direct band-gap at 18 GPa in Fig.A2 (a) above, this shows that the tube remains semi-conducting at this pressure. As the pressure is increased the tube becomes metallic and we observe this as crossing of the lower states of the conduction band at pressure of 21 GPa in Fig.A2 (b). The sample band structure at 30 GPa in Fig.A2 (c) shows further splitting of the states close to the Fermi level in the conduction as well as in the valence band with tube showing metallic behavior. At higher pressures the states moves such that a direct band-gap is formed at the Fermi level as in Fig. A2 (d). The tube at this pressure shows semi-conducting behavior and is also observed in the 1dimensional plot in Fig. 2(d)

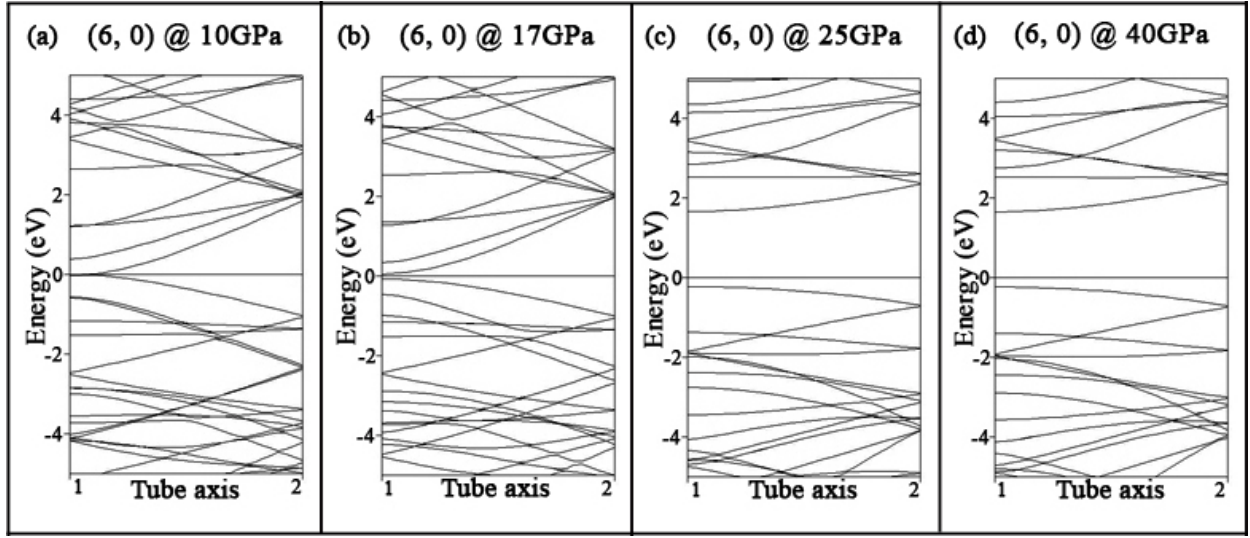

Fig.A3: - The plot of band structure diagram for metallic (6, 0) SWCNT are shown above in Fig. (a) – Fig.(d) at pressures of 10 GPa, 17 GPa, 25 GPa and 40 GPa. The Fermi level is at 0 eV.

The band structure diagram at 10 GPa in Fig.A3 (a) show the crossing of the states in the valence and conduction band at the Fermi level. As the pressure further increases, the overlap of the states decrease as a result of which the DOS at the Fermi level decreases as observed in the 1-dimensional inset in Fig 3(d). At higher pressures the tubes come so close together that they start to interact and deform due to interaction. Due to this a large band gap opens up as observed in Fig.A3 (c) and remains almost constant at higher pressures as sampled in Fig.A3 (d) above. This can also be seen as a black patch in the pressure range 25GPa – 60GPa in the DOS contour plot in Fig 3(d).

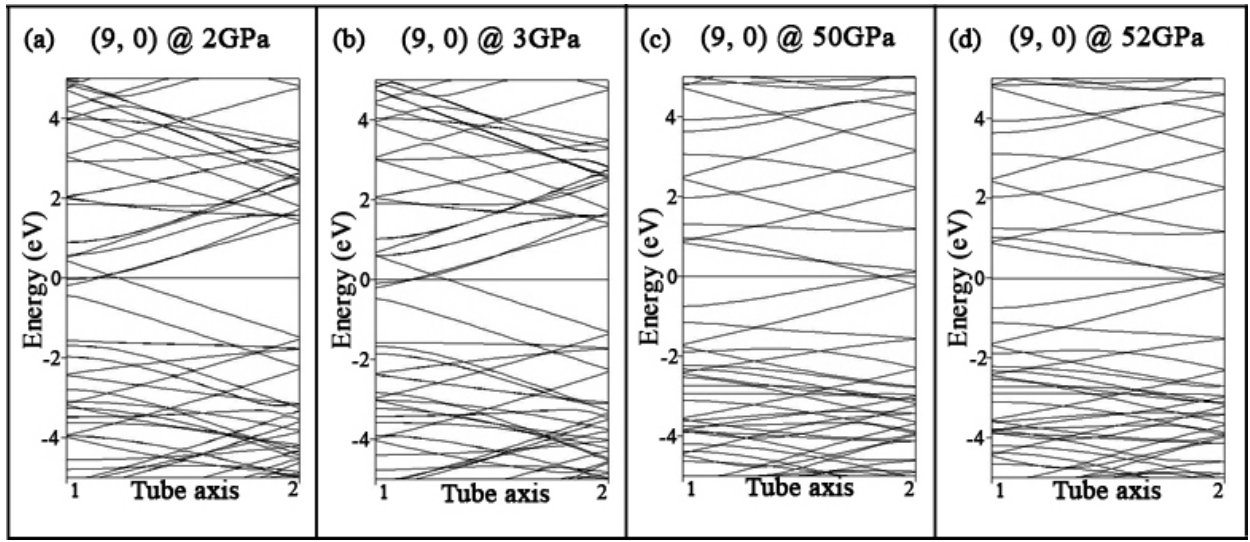

Fig.A4: - The plot of band structure diagram for metallic (6, 0) SWCNT are shown above in Fig. (a) – Fig.(d) at pressures of 2 GPa, 3 GPa, 50 GPa and 52 GPa. The Fermi level is at 0 eV.

Fig.A4 (a) and Fig.A4 (b) above show overlapping of the conduction and the valence bands indicating the metallic behavior. As the pressure is increased from 2 GPa to 3 GPa splitting of the lowest conduction and valence band occurs. At higher pressure around 50 GPa as sampled in Fig.A4 (c) and Fig. A4 (d) above we observe that due to crossing of both the conduction as well as the valence bands the tube remains metallic.

**FIGURE CAPTIONS**

Figure 1: - Charge density plots for the (7, 0) nanotubes in a supercell showing 4 unit cells at pressures of (a) 8 GPa, (b) 20 GPa and (c) 60 GPa. The white arrows indicate the bonds interlinking two tubes while black arrows are bonds within the same tube. (d) Contour plot of DOS for pressures up to 60 GPa. The inset at the top shows the variation of the DOS at the Fermi level for all pressures. The pressures corresponding to panels (a), (b) and (c) are indicated as arrows. The lower right figure shows the semi-conducting sample DOS plot at 20 GPa.

Figure 2: - Charge density plots for the (8, 0) tubes in a supercell showing 4 unit cell at pressures of (a) 8 GPa, (b) 30 GPa and (c) 60 GPa. The white arrows marked refer to the weaker bonds while Black arrows refer to normal C-C bonds in the nano tube. (d) Contour plot in the pressure range upto 60 GPa. The inset at the top shows the variation of the DOS at the Fermi level for all pressures. The pressures corresponding to panels (a), (b) and (c) are indicated as arrows. The lower right figure shows the metallic sample DOS plot at 30 GPa.

Figure 3: - Charge density plots for the (6, 0) tubes in a supercell showing 4 unit cell at pressures of (a) 8 GPa, (b) 20 GPa and (c) 60 GPa. The white arrows refer to the bonds interlinking two tubes (d) Contour plot in the pressure range upto 60 GPa. The inset at the top shows the variation of the DOS at the Fermi level for all pressures. The pressures corresponding to panels (a), (b) and (c) are indicated as arrows. The lower right figure shows the metallic sample DOS plot at 20 GPa.

Figure 4: - Charge density plots for the (9, 0) tubes in a supercell showing 4 unit cell at pressures of (a) 8 GPa, (b) 30 GPa and (c) 60 GPa. The white arrows marked refer to the representative weaker interlinking bonds while Black arrows refer to normal C-C bonds in the nanotube. (d) Contour plot in the pressure range upto 60 GPa. The inset at the top shows the variation of the DOS at the Fermi level for all pressures. The pressures corresponding to panels (a), (b) and (c) are indicated as arrows. The lower right figure shows the metallic sample DOS plot at 30 GPa.

**Fig. 1. Sumit and Tyson**

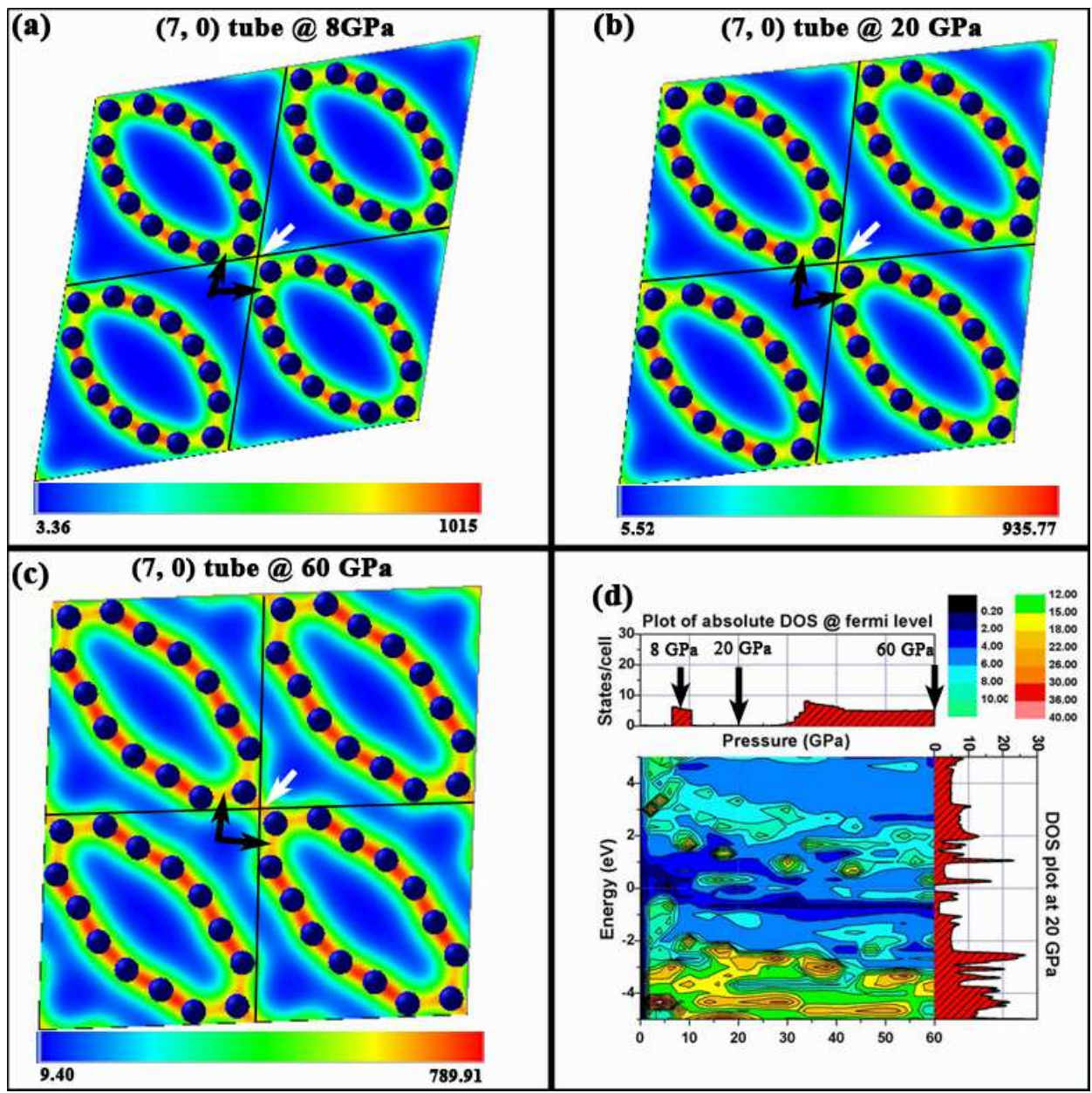

**Fig. 2. Sumit and Tyson**

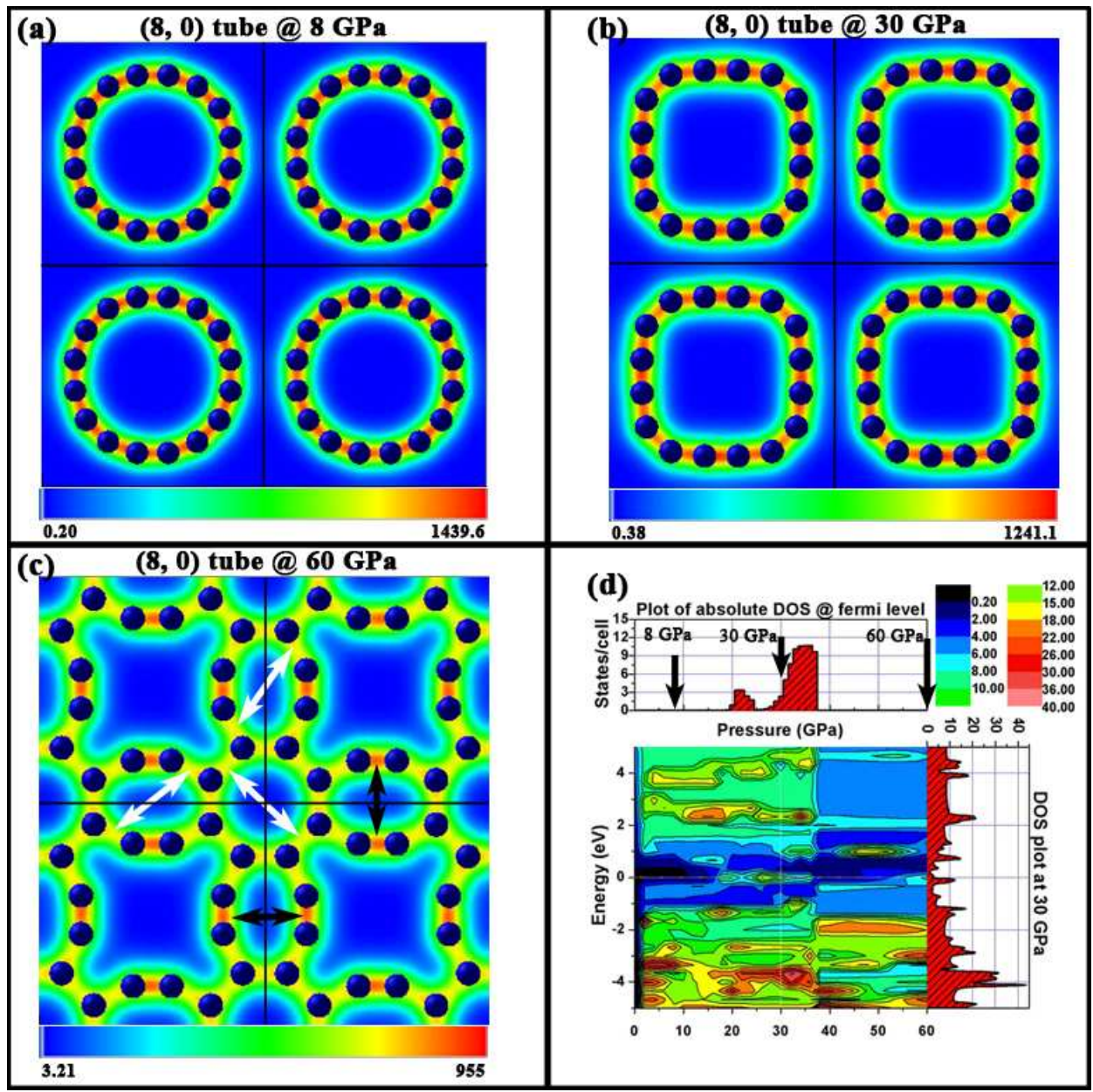

**Fig. 3. Sumit and Tyson**

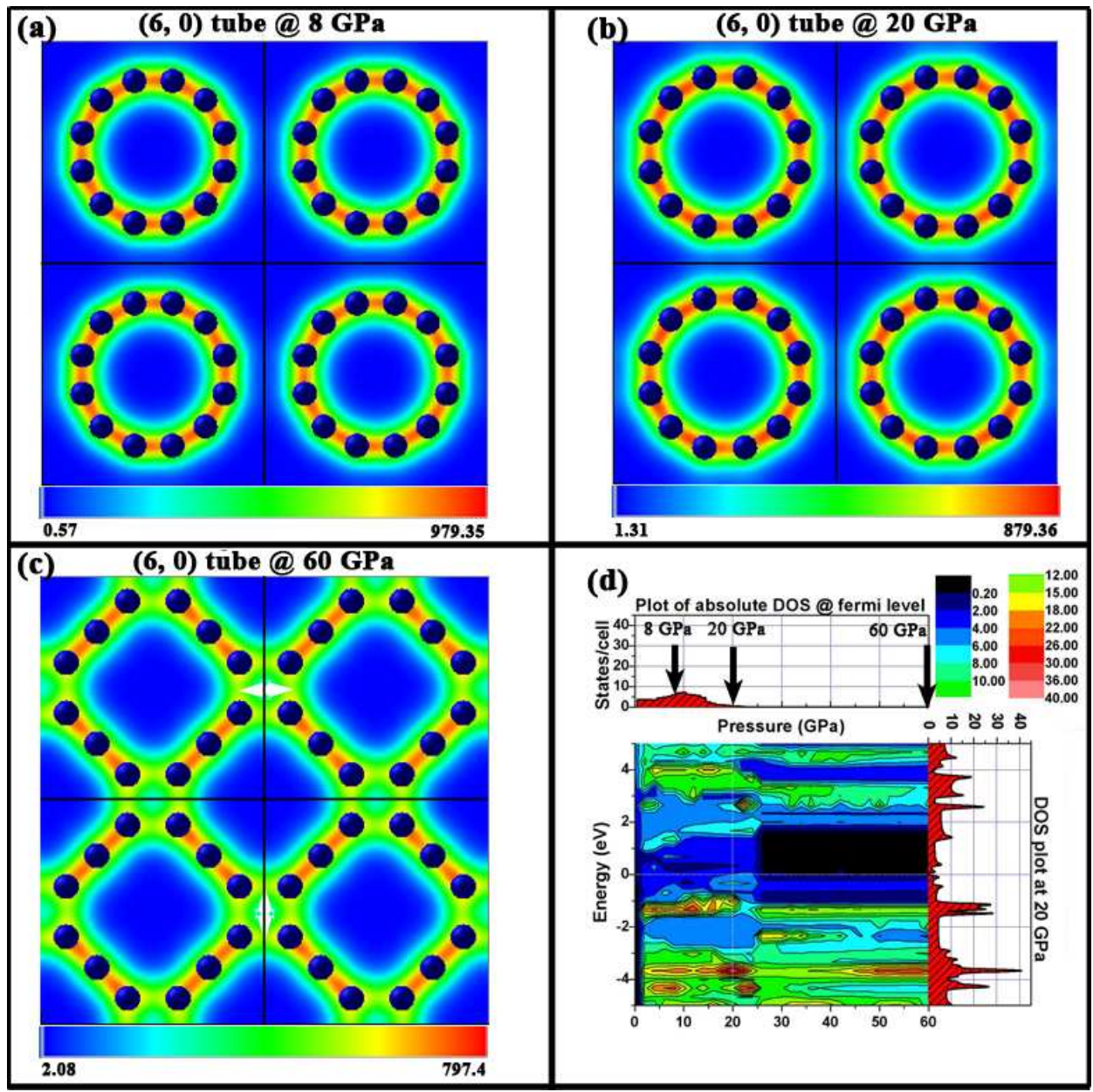

**Fig. 4. Sumit and Tyson**

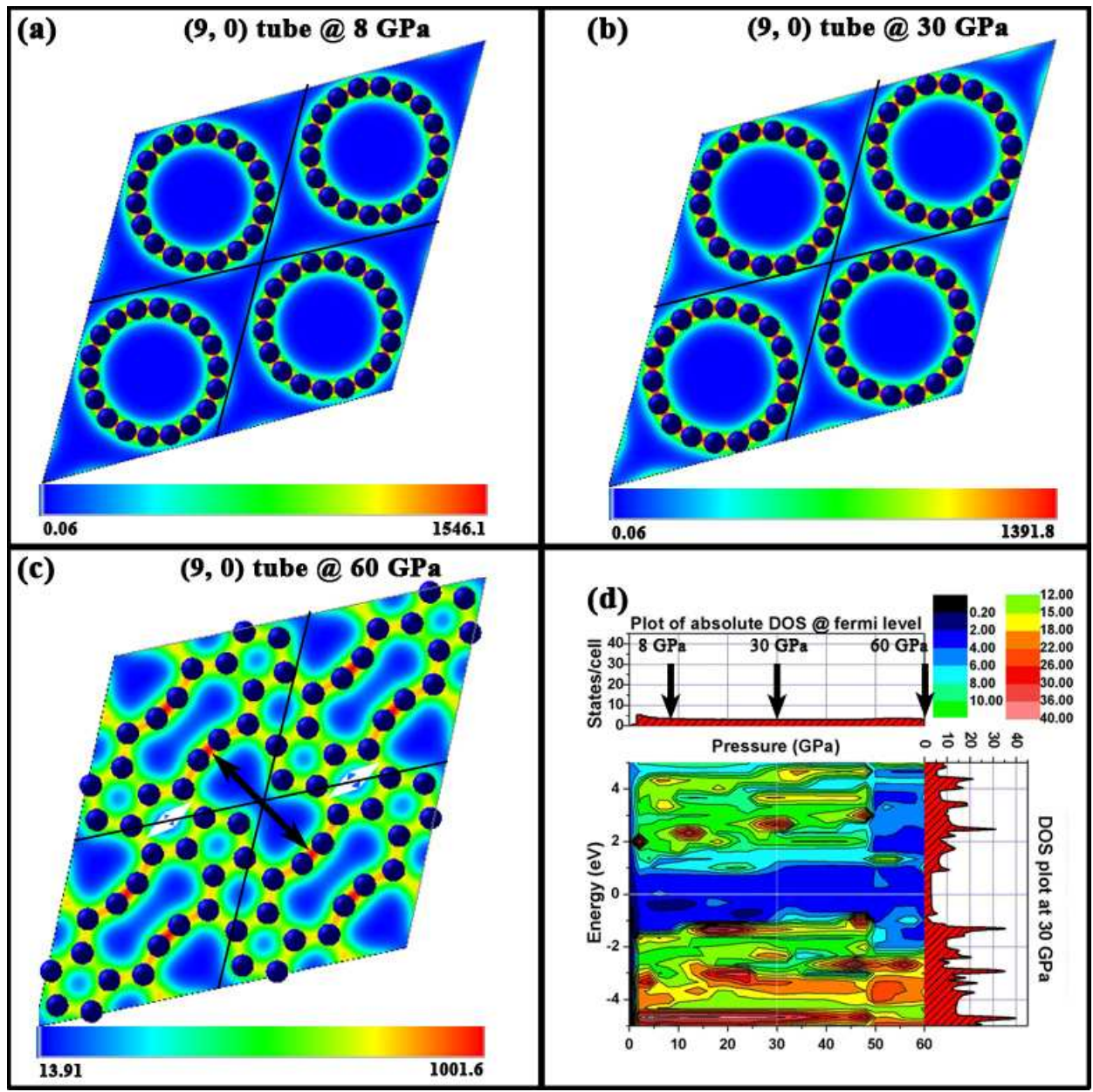

**References: -**

[1] R. H. Baughman, A. A. Zakhidov, and W. A. De Heer, Science 297, 787 (2002).
[2] H. W. C. Postma, T. Teepen, Z. Yao, M. Grifoni, and C. Dekker, Science 293, 76 (2001).
[3] P. Gopinath, A. Mohite, H. Shah, J. T. Lin, and B. W. Alphenaar, Nano Letters 7, 3092 (2007).
[4] T. W. Tombler, C. Zhou, L. Alexseyev, J. Kong, H. Dai, L. Liu, C. S. Jayanthi, M. Tang, and S. Y. Wu, Nature 405, 769 (2000).
[5] L. Vitali, M. Burghard, P. Wahl, M. A. Schneider, and K. Kern, Physical Review Letters 96, 1 (2006).
[6] R. C. Stampfer, T. Helbling, D. Obergfell, B. Scho?berle, M. K. Tripp, A. Jungen, S. Roth, V. M. Bright, and C. Hierold, Nano Letters 6, 233 (2006).
[7] A. Maiti, Nat Mater 2, 440 (2003).
[8] J. Zang, A. Treibergs, Y. Han, and F. Liu, Physical Review Letters 92, 105501 (2004).
[9] E. D. Minot, Y. Yaish, V. Sazonova, J. Y. Park, M. Brink, and P. L. McEuen, Physical Review Letters 90 (2003).
[10] J. Tang, L. C. Qin, T. Sasaki, M. Yudasaka, A. Matsushita, and S. Iijima, Physical Review Letters 85, 1887 (2000).
[11] L. Yang and J. Han, Physical Review Letters 85, 154 (2000).
[12] P. E. Lammert, P. Zhang, and V. H. Crespi, Physical Review Letters 84, 2453 (2000).
[13] M. R. Falvo, G. J. Clary, R. M. Taylor Ii, V. Chi, F. P. Brooks Jr, S. Washburn, and R. Superfine, Nature 389, 582 (1997).
[14] S. W. Lee, G. H. Jeong, and E. E. B. Campbell, Nano Letters 7, 2590 (2007).
[15] G. Kresse and J. Furthmuller, Computational Materials Science 6, 15 (1996).
[16] G. Kresse and J. Furthmuller, Physical Review B - Condensed Matter and Materials Physics 54, 11169 (1996).
[17] G. Kresse and D. Joubert, Physical Review B 59, 1758 (1999).
[18] J. P. Perdew, K. Burke, and M. Ernzerhof, Physical Review Letters 77, 3865 (1996).
[19] O. Dubay, G. Kresse, and H. Kuzmany, Physical Review Letters 88, 235506 (2002).
[20] O. Dubay and G. Kresse, Physical Review B - Condensed Matter and Materials Physics 70, 1 (2004).
[21] O. Dubay and G. Kresse, Carbon 42, 979 (2004).
[22] J. Wu, et al., Physical Review Letters 93, 017404 (2004).
[23] E. Durgun, S. Dag, V. M. K. Bagci, O. Gu?lseren, T. Yildirim, and S. Ciraci, Physical Review B - Condensed Matter and Materials Physics 67, 2014011 (2003).